# Low kinetic inductance superconducting MgB$_2$ nanowire photon detectors with a 100 picosecond relaxation time.

*Sergey Cherednichenko\*, Narendra Acharya, Evgenii Novoselov, and Vladimir Drakinskiy*

Terahertz and Millimetre Wave Laboratory, Department of Microtechnology and Nanoscience, Chalmers University of Technology, SE-412 96, Gothenburg, Sweden

*\* Corresponding author serguei@chalmers.se*

**Abstract**. Properties of superconducting nanowires set the performance level for Superconducting Nanowire Single Photon Detectors (SNSPD). Reset time in commonly employed large area SNSPDs, 5-10ns, is known to be limited by the nanowire's kinetic inductance. On the other hand, reduction of the kinetic inductance in small area (waveguide integrated) SNSPDs prevents biasing them close to the critical current due to latching into a permanent resistive state. In order to reduce the reset time in SNSPDs, superconducting nanowires with both low kinetic inductance and fast electron energy relaxation are required. In this paper, we report on narrow (15-100nm) and long (up to 120μm) superconducting MgB$_2$ nanowires offering such combination of properties. In 5 nm-thick MgB$_2$ films, grown using Hybrid Physical Chemical Vapor Deposition, the electron relaxation time was 12ps, the critical temperature was ~32K, and the critical current density was ~$5\times10^7$A/cm$^2$ (at 4.8K). Using microwave reflectometry, we measured a kinetic inductance of $L_{k0}$(4.8K)=1.3-1.6 pH/□ regardless of the nanowire width, which results in a magnetic field penetration depth of ~90 nm. These values are very close to those in pristine MgB$_2$. For 120 μm long nanowires the response time was only ~100ps, i.e. 1/80 of that in previously reported NbN nanowire photon detectors of similar dimensions.

**Main text.**

During the last two decades we have seen enormous progress with Superconducting Nanowire Single Photon Detectors (SNSPD), from first demonstration of priciples[1] to integrated circuits and a wide range of applications in nanophysics[2], high speed communication[3], quantum information[4], quantum key distribution[5], laser-ranging[6], deep space communication[7], etc. SNSPD's operation is based on photon assisted disruption of supercurrent in a superconducting nanowire, which is dc- biased close to the critical current.[8,9] Because the superconducting energy gap $\Delta$ is in the meV-range, a large number of non-equilibrium quasiparticles are generated at visible or infrared photon (~eV) impacts. Both the small width of the nanowire, $w$ and the critical current, $I_c$ being close to the pair-breaking current ensure high detection efficiency. [10,11] After hot-spot formation, the bias current is diverted from the SNSPD into the readout. Within a relaxation time, $\tau_0$ (set by electron-phonon non-elastic scattering and phonon escape into the substrate), non-equilibrium quasiparticles recombine into Cooper pairs. Until the supercurrent in the nanowire is restored and the hot-spot disappears (the reset time), SNSPD is blind to any further incoming photons. In 5nm thick NbN films, the quasiparticle relaxation time $\tau_0$ is ~50ps[12]. However, in long ($l$) and narrow ($w$)

NbN nanowires, large kinetic inductivity ($L_{k0}(4.8K)$=90 pH/□) (inductance per square of the nanowire, $l=w$) leads to a significantly slower current return (restoration of the superconducting state), resulting in a much larger reset time $\tau$ [13] : $\tau=L_k/R_L >> \tau_0$, where $L_k = L_{k0} \times l/w$ is the total kinetic inductance of the nanowire and $R_L$ is the impedance of the read-out circuit (~50 Ω). This is particularly critical in large area SNSPDs, with a kinetic inductance as large as hundreds of nH, leading to a reset time of a few tens of ns. On the other hand, large kinetic inductance delays current return from the load $R_L$ back into the SNSPD, prior to the hot-spot cooling down, hence preventing the SNSPD from latching into a permanent normal (non-sensitive) state.[14] To a certain extent, the reset time can be reduced by choosing a high-impedance load. However, much increased $R_L$ leads to a reduction of the maximum non-latching bias currents, hence precluding SNSPDs from reaching high detection efficiencies.[15] This problem is particularly pronounced in short SNSPD, where shunting resistors of small values or extra inductors in series are required to avoid latching, which does not altogether allow reaching short reset times. [16,17]

Apart from low-$T_c$ superconductors (NbN, Nb, NbTiN, TaN, MoN, WSi, etc.) [10, 14, 18, 19], magnesium diboride (MgB$_2$) nanowires have also shown ability for single photon detection in both visible[20] and IR[21] ranges. However, the reset time in MgB$_2$ SNSPDs was reported to be 2-3 ns, i.e. close to that for NbN SNSPDs. Previously it has been shown that in clean MgB$_2$ (attainable in 50-200nm thick films) the magnetic field penetration depth $\lambda$ is at least an order of magnitude shorter than that in NbN,[22] which should lead to a small kinetic inductivity ($L_{k0}=\mu_0\lambda^2/d$). Moreover, we recently reported that 5nm-thick MgB$_2$ films could be made with a critical temperature >30K and a normal state resistivity comparable to that in thick films. In such films, a quasiparticle relaxation time is $\tau_0$=12ps [23], and which all together suggests that MgB$_2$ nanowires might provide a solution for high speed photon detection.

In this work, we report on the successful demonstration of $MgB_2$ nanowires made from clean and ultra- thin (5nm) films. We study both kinetic inductance and response rate in the devices within a large variety of widths (15 nm – 900 nm) and lengths (3 μm – 120 μm), and show that a 100 ps voltage relaxation time can be achieved in 120 μm- long devices with a standard 50 Ω readout without latching.

$MgB_2$ films were grown on 6H-SiC substrates using a custom-built Hybrid Physical Chemical Vapor Deposition (HPCVD) system.[24,25] The system was designed for a low deposition rate (~2.5-3 nm/min) with a diborane gas flow of 2 sscm (5 % $B_2H_6$ in $H_2$) and a deposition temperature of 700°C. As a result, continuous $MgB_2$ films could be routinely obtained as thin as 5 nm, as shown by Transmission Electron Microscopy (TEM) (Fig.1a), with a critical temperature of 32-34 K (Fig.3a). Film growth on the SiC substrate is epitaxial, which ensures that bulk properties of $MgB_2$ are preserved in the thin films as manifested in a fairly low normal-state (residual) resistivity of $\rho_n$=10-15 μΩ×cm, a high critical current density of $J_c$~5×10$^7$A/cm$^2$, and a small magnetic field penetration depth of $\lambda$=90nm (see further in the text). Nanowires (Fig.1 b,c) were fabricated using e-beam lithography and $Ar^+$ ion milling through a negative resist mask.[26] Nanowires were from ~15 nm to 900 nm in width (determined from Scanning Electron Microscope (SEM) images), and from 3 μm up to 120 μm in length. All experiments were conducted in a cryogen-free probe station (base temperature 4.8 K) with an optical view port and microwave ground- signal- ground (GSG) probes (Fig.2). Kinetic inductance was obtained from the imaginary part of the microwave impedance ($Z(\omega)=j\omega L_k$), which was measured using a (10MHz-67GHz) Vector Network Analyzer. Voltage response was registered with a real-time digital oscilloscope Infiniium 54854, with nanowires being dc-voltage-biased through a set of low-pass filters and a bias-T (see also Supplementary Material).

As measured in a large variety of samples (width and length), the kinetic inductivity in thin MgB$_2$ nanowires was 1.35-1.60 pH/□ at 4.8K (Fig.3b). In the case of defect-free uniform films, the total kinetic inductance $L_k = L_{k0} \frac{l}{w} = L_0 \frac{R}{R_s}$ is expected to be proportional to the length-to-width ratio ($l/w$) and hence to the total resistance $R$ in the normal state (40K). This trend was indeed observed (Fig.3c), which confirms scalability of thin-film transport properties from the micron- to the nano- scales.

Kinetic inductance (and its temperature dependence) is a sensitive indicator for superfluid (Cooper pairs) density, hence providing the value for the superconducting energy gap. In MgB$_2$, π- and σ-sheets of the Fermi surface display quite dissimilar energy gaps, with temperature dependences following:[27]

$$\Delta_\pi(T) = \Delta_\pi(0) \cdot \left[1 - \left(\frac{T}{T_c}\right)^{1.8}\right]^{\frac{1}{2}} \; ; \; \Delta_\sigma(T) = \Delta_\sigma(0) \cdot \left[1 - \left(\frac{T}{T_c}\right)^{2.9}\right]^{\frac{1}{2}} \qquad (1)$$

Previously, in the frame of double-band modeling[28], it has been shown that electrodynamic properties in MgB$_2$ can be analyzed by considering two quasi-layers (electrons of both π- and σ- bands) in parallel, with a finite (yet, weak) interband scattering. In particular, partial contribution from each band to both the condensate energy and the superfluid density varies according to temperature and magnetic field. E.g. at zero magnetic field, the π-band is expected to contribute about $a$>70% to the superfluid density ($\propto 1/\lambda^2$) at $T$<($T_c$-2K). The exact value of the coefficient $a$ depends on the interband scattering rate, and has to be evaluated for each particular case. However, the general trend seems to be correct, as in Ref.[22], where it was shown that $\lambda(0)$ vs the mean free path can be modeled considering predominantly the π-band. In Ref.[29], a satisfactory agreement with experimental $\lambda(T)$ was demonstrated for thick films, considering two bands with an interband coefficient a= 0.8. For a general case of two conduction bands, the kinetic inductivity can be calculated from the imaginary part of the BCS conductivity[30] as:

$$L_{k0}(T) = \left[\left(\frac{\hbar \cdot 2R_S}{a \cdot \pi \cdot \Delta_\pi(T)} \cdot \frac{1}{tanh\left(\frac{\Delta_\pi(T)}{2k_B \cdot T}\right)}\right)^{-1} + \left(\frac{\hbar \cdot 2R_S}{(1-a) \cdot \pi \cdot \Delta_\sigma(T)} \cdot \frac{1}{tanh\left(\frac{\Delta_\sigma(T)}{2k_B \cdot T}\right)}\right)^{-1}\right]^{-1} \quad (2)$$

By fitting Eq (2) to the experimental $L_{k0}$(T) data (Fig. 4a), we obtained the superconductor energy gaps at $T$=0. Our results showed that Eq (1) holds even for extremely thin MgB$_2$ films considering that $\Delta_\pi(0) = 0.6 k_B \cdot T_c = 1.7\ meV$ and $\Delta_\sigma(0) = 2.2\ k_B \cdot T_c = 6.2\ meV$, with an interband coefficient $a$=0.85. For a mean experimental value of $L_{k0}$(5K)=1.5pH/□, we can obtain the effective magnetic field penetration depth: $\lambda_{eff}$(5K)= $(L_{k0} \times d / \mu_0)^{1/2}$ = 90nm.

In order to study the effect of kinetic inductance on the voltage response in MgB$_2$ nanowires, we recorded voltage pulses (Fig.4b) generated by either spontaneous (dark) or photon induced (from a λ=630nm cw laser) excitations, while nanowires were biased close to the corresponding critical currents.

For the longest (in the number of squares) samples (30nm×120μm, 3870 squares), the measured inductance was 5153pH, which should result in a voltage fall time of $\tau=L_k/R_L$=103ps for a 50 Ω readout. This is close to the value which we found experimentally (Fig.4b). With some ringing on the voltage decay section, caused by a sharp out-of-bandwidth roll-off in the oscilloscope, the voltage fall time (at the 1/e- level) is ~100 ps for the longest samples. At this time scale, the intrinsic quasiparticle energy relaxation time (12ps) can be neglected, whereas it will be the limiting factor in short nanowires. However, the fall time in shorter samples was only marginally shorter than in long samples, which is a consequence of a readout bandwidth limitation imposed both by the amplifier and the oscilloscope (~2 GHz, see Supplementary Material). For comparison, a 100nm-wide NbN SNSPD which covers an area of 10μm×10μm with a filling factor of 50% (~500 μm long) has a similar $L/w$ ratio to our longest MgB$_2$ samples. For such NbN devices, a voltage fall time of ~8ns has previously been reported [13], i.e. a factor of 80 longer than that in MgB$_2$.

Fast quasiparticles relaxation (12 ps in MgB$_2$ vs 50ps in NbN) is a positive factor and reduces the effect of latching in small kinetic inductance samples. We were able to register multiple photon detection events (i.e. no lathing took place) for short MgB$_2$ samples biased directly at the critical current. By increasing laser intensity, the number of pulses on the oscilloscope increases yet preserves fast voltage relaxation.

In conclusion, our results show that fabrication of narrow and long nanowires is feasible using MgB$_2$ superconducting ultra-thin films made by a low deposition rate HPCVD process. With the preserved high quality of films manifested in a high critical temperature (>30K), a high critical current density (~5×10$^7$A/cm$^2$) and a small magnetic penetration depth (~90 nm), MgB$_2$ nanowires have a high potential for fast response rate photon (as well as particle) detection, particularly for large area detectors. With response time being limited by kinetic inductance, MgB$_2$ nanowire detectors are expected to show a reset factor of $L_{k0}$(NbN)/$L_{k0}$(MgB$_2$)=90pH/□ / 1.5pH/□ =60 times faster compared to NbN devices with similar $L/w$ ratios. In shorter MgB$_2$ detectors, the observed voltage relaxation time (~90ps) remains limited by the utilized readout bandwidth (2GHz). However, the questions of how short MgB$_2$ nanowires can be, while avoiding the problem of latching, and whether presented MgB$_2$ nanowire photon detectors are capable of single photon sensitivity, are still to be investigated.

ACKNOWLEDGMENT

The authors would like to thank Dr. Olof Bäcke for conducting TEM analysis of MgB$_2$ samples; Professors X. X. Xi and K. Chen for sharing their experience with the HPCVD process for MgB$_2$ films and enlightening discussions; and Usman Ul-Haq for the help with process development. MgB$_2$ film technology was developed under an ERC grant 308130-Teramix. This research was also supported by Swedish Research Council (grant # 2016-04198) and by Swedish National Space Agency (grant # 198/16).

SUPPLEMENTARY MATERIAL

See supplementary material for details on MgB$_2$ film deposition, MgB$_2$ nanowire fabrication, experimental set-up and examples for measured microwave impedances of MgB$_2$ nanowires.

Figures.

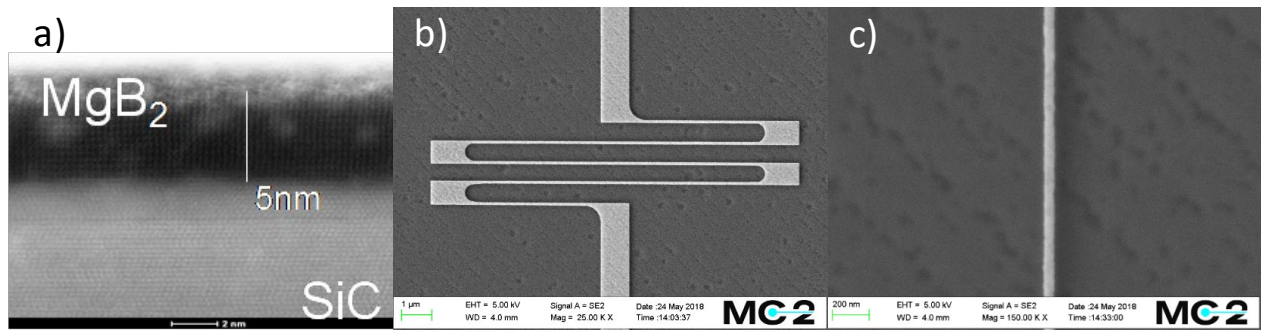

Fig. 1 (a) TEM image of a 5nm thick MgB$_2$ film on the SiC substrate. SEM images of a meandering (b) and a straight (c) MgB$_2$ nanowires. Scale bars are given in each of the images.

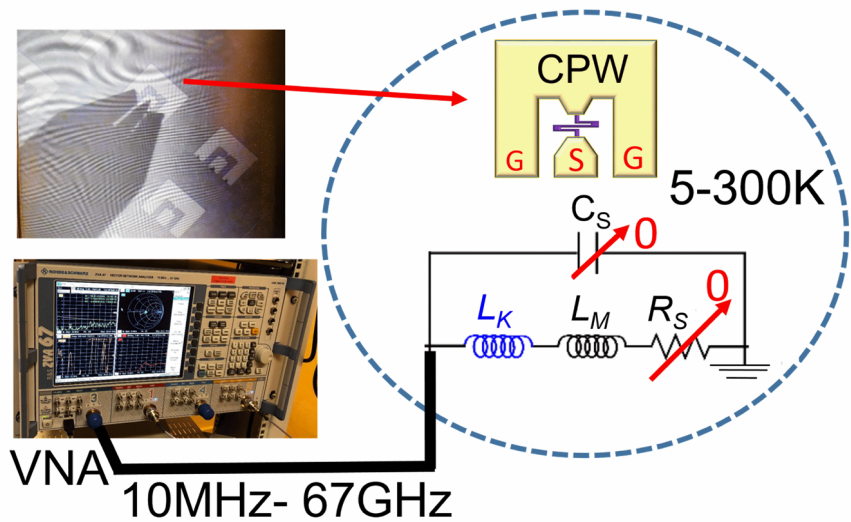

Fig. 2 The kinetic inductance measurements set-up. Chip layout (right), samples on the 5K stage in the cryo-probe station with the microwave probe contacted (the view through the pressure window and the IR filter) (left-top), the Vector Network Analyzer (left-bottom).

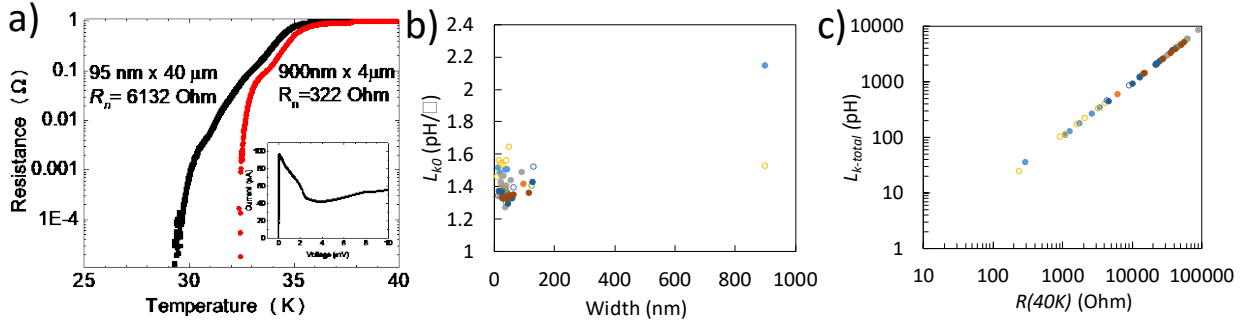

Fig. 3 (a) Superconducting transition in 95nm- and 900nm- wide nanowires. Inset: an *I(V)* curve for a 40nm×120μm MgB$_2$ nanowire. (b) Kinetic inductivity (at 4.8 K) for all samples with various widths and lengths. (c) The total kinetic inductance (at 4.8 K) vs the normal state resistance *R*(40K) for all samples.

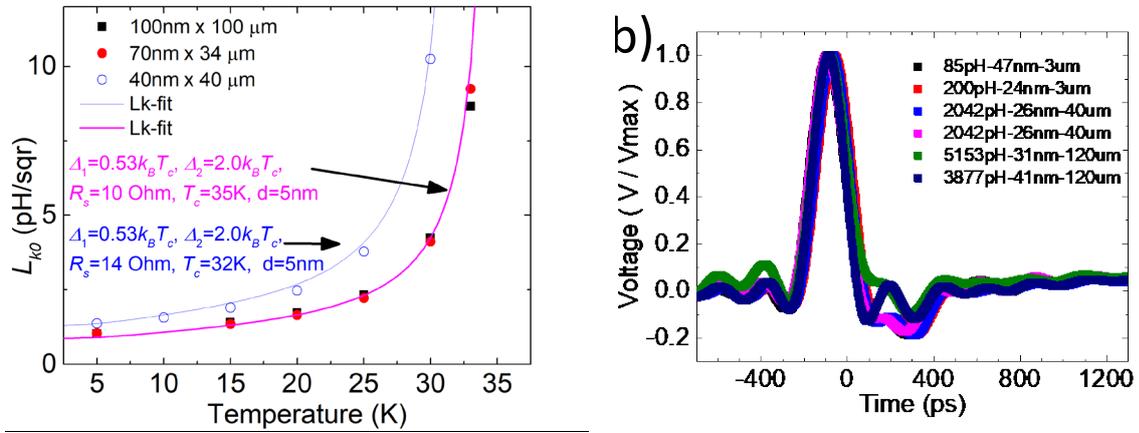

Fig. 4 (a) Kinetic inductivity vs temperature for nanowires from two batches with critical temperatures of 35K and 32.7K. The fits are made using equation (2), where energy gaps for both π- and σ- bands were calculated using equation (1) and the experimental T$_c$. b) Voltage response of MgB$_2$ nanowire photon detectors with various total kinetic inductances (width and length variation).